\documentclass[useAMS]{mn2e}
\usepackage{psfig}

\usepackage{times}
\usepackage{amssymb,amsmath}

\def\lsim{ \lower .75ex\hbox{$\sim$} \llap{\raise .27ex \hbox{$<$}} }
\def\gsim{ \lower .75ex \hbox{$\sim$} \llap{\raise .27ex \hbox{$>$}} }

\title[Extreme blazars and IGMF] 
{Extreme TeV blazars and the intergalactic magnetic field}

\author[Tavecchio et al.]
{F. Tavecchio$^1$\thanks{E--mail: fabrizio.tavecchio@brera.inaf.it},
G. Ghisellini$^1$, G. Bonnoli$^1$ , L. Foschini$^1$ \\
$^1$INAF -- Osservatorio Astronomico di Brera, via E. Bianchi 46, I--23807
Merate, Italy\\
}

\begin{document}



\maketitle

\begin{abstract} 
We study the four BL Lac objects (RGB J0152+017, 1ES 0229+200, 1ES 0347-121 and PKS 0548-322) detected in the TeV band but not present in the 1FGL catalogue of the {\it Fermi}/Large Area Telescope. We analize the 24 months of LAT data deriving $\gamma$--ray fluxes or upper limits that we use to assemble their spectral energy distributions (SED). We model the SEDs with a standard one-zone leptonic model, also including the contribution of the reprocessed radiation in the multi GeV band, emitted by the pairs produced through the conversion of the primary TeV emission by interaction with the cosmic optical-IR background. For simplicity, in the calculation of this component we adopt an analytical approach including some simplifying assumptions, in particular i) the blazar high energy emission is considered on average stable over times of the order of $10^{7}$ years and ii) the observer is exactly on-axis. We compare the physical parameters derived by the emission model with those of other high-energy emitting BL Lacs, confirming that TeV BL Lacs with a rather small GeV flux are characterized by extremely low values of the magnetic field and large values of the electron energies. The comparison between the flux in the GeV band and that expected from the reprocessed TeV emission allows us to confirm and strengthen the lower limit of $B \gtrsim 10^{-15}$ G for the intergalactic magnetic field using a theoretically motivated spectrum for the primary high-energy photons.
\end{abstract}
 

\section{Introduction}

Blazars, characterized by a relativistic jet pointing towards the Earth, represent the most extreme flavor of Active Galactic Nuclei (AGN). Their relativistically-boosted non-thermal continuum, extending from radio to $\gamma$--ray energies, is characterized by a spectral energy distribution (SED) with two broad peaks, with maxima at IR/UV or even X-ray band (due to synchrotron emission by relativistic electrons in the jet) and in the $\gamma$--ray band (probably produced through inverse Compton scattering), respectively. Fossati et al. (1998) proposed the existence of an inverse relation between the frequencies of both peaks and the (radio or bolometric) luminosity, the so called ``blazar sequence".  The most powerful sources are thus characterized by peaks located at low frequencies, while the less powerful ones display the highest peak frequencies (hence the classification as Highly peaked BL Lac objects, HBLs, Padovani \& Giommi 1995) and their high-energy bump can reach TeV energies. Sources detected at Very High Energies (VHE, $E>100$ GeV) still form a  small population\footnote{see \tt http://www.mpp.mpg.de/$\sim$rwagner/sources/} by they are intensively studied (e.g. De Angelis et al. 2008), since they represent ideal natural laboratories to investigate particle acceleration and cooling and to indirectly probe the extragalactic background light (EBL). 

Tavecchio et al. (2010a, hereafter T10a), studying all the BL Lac of the LBAS sample (Abdo et al. 2009; it includes all the AGN detected at high significance by the Large  Area Telescope onboard {\it Fermi} in the first three months of operation), and/or at VHE by  Cherenkov telescopes, discussed the possible existence of two separate groups of high-energy detected BL Lacs. In fact, while the majority of them, when fitted with a standard one-zone synchrotron self Compton model, are characterized by values of the magnetic field in the range between 0.1-1 G and Lorentz factors of the electrons emitting at the peak of the order of $\gamma _{\rm b}\sim 10^4-10^5$, a handful of sources are characterized by very low magnetic fields (of the order of $10^{-2}-10^{-3}$ G)  and very large electron energies, corresponding to Lorentz factors of  $\gamma _{\rm b}\sim 10^6-10^7$. These peculiar values of the physical quantities are directly connected to the extremely high frequencies of the peaks in the SED (hence the name of ``extreme" HBLs introduced by Costamante et al. 2001).

Because of the extremely low flux characterizing these sources in the GeV band, some of these extreme HBLs have been recently considered for deriving upper limits on the still unknown intensity of the intergalactic magnetic field (IGMF, e.g. Kronberg 2001, Widrow 2002), based on the measured upper limits of the flux in the GeV band (Neronov \& Vovk 2010, Tavecchio et al. 2010b, hereafter T10b, Dolag et al. 2011). 

Constraints on the magnetic field intensity in voids could help to understand the origin of the ``seed" fields assumed in dynamo amplification models for magnetic fields in galaxies and galaxy clusters (e.g. Kulsrud \& Zweibel 2008). Proposed models for these seed fields range from those considering effects occurring during the inflation (e.g. Turner \& Widrow 1988) or phase transition era in the Early  Universe (e.g. Kahniashvili et al. 2011 and references therein) to those invoking mechanisms active during the early stages of protogalaxy formation (e.g. Gnedin et al. 2000). Classical methods used to measure or derive {\it upper} limits on IGMF consider Faraday rotation of polarization angle of radio emission of quasars (e.g. Kronberg 2001) or the effects of magnetic fields on the Cosmic Microwave Background (e.g. Durrer et al. 2000). 

A rather promising method to obtain {\it lower limits} to the field (hence complementary to the methods discussed above) is based on the fact that VHE photons emitted by blazars are absorbed and converted into electron-positron pairs through the interaction with the optical-IR cosmic background (e.g. Nikishov 1962). These pairs, in turn, rapidly cool through inverse Compton scattering off photons of the CMB, emitting photons of lower energies. If the primary spectrum extends up to very high energies, part of the reprocessed emission will still be above the threshold for absorption and a further generation of pairs will be created. If the maximum energy of the primary emission is sufficiently large this process will lead to the formation of an electromagnetic cascade (e.g. Aharonian et al. 1994, Coppi \& Aharonian 1997). While the pairs emit, they are deviated by the original trajectory by the tiny IGMF.  Therefore, the observed reprocessed emission will be spread over a solid angle larger than the original one, ``diluting" its flux: the comparison between the predicted and the observed reprocessed flux provides a direct measure or a lower limit of the IGMF intensity (Plaga 1995, Dai et al. 1992, D'Avezac et al. 2007, Murase et al. 2008, Dolag et al. 2009, Elyv et al. 2009). 

In this paper we consider the extreme HBLs not detected by LAT in the first 11 months of observations (1FGL catalogue). 
In this catalogue four TeV HBLs are not present, namely  RGB J0152+017 ($z=0.080$), 1ES 0229+200 ($z=0.14$), 1ES 0347-121 ($z=0.188$) and PKS 0548-322 ($z=0.069$). 

The  aim of this paper is twofold: 1) we intend to model the SED of the sources and derive the physical parameters considering also the possible role of the reprocessed emission in shaping the observed high-energy spectrum; 2) we use the modeling of the primary blazar high-energy emission to calculate self-consistently the reprocessed component and therefore derive the lower limits on the IGMF. We stress that both issues are necessarily interlaced: the reprocessed component can provide a non-negligible contribution to the 10-100 GeV emission (or even above), while the primary TeV spectrum is necessary to have a reliable characterization of the expected reprocessed component.

For all the sources we analize the $\approx$24 months LAT data deriving upper limits (or fluxes, in few cases) in the 0.1-100 GeV band that we use, together with data at other frequencies, to construct their SEDs (\S 2). We model their SEDs with the standard one-zone leptonic model and we derive the physical parameters of the emitting region (\S 3) also considering the possible contribution of the IC emission from the pairs produced by the conversion of high-energy photons interacting with the EBL,  whose level and spectral shape also depends on the value of the IGMF. In \S 4 we show the results and in \S 5 we conclude.

We use $H_{\rm 0}\rm =70\; km\; s^{-1}\; Mpc^{-1} $, $\Omega_{\Lambda}=0.7$, $\Omega_{\rm M} = 0.3$.

\section{Spectral Energy Distributions}

\subsection{LAT data}
RGB J0152+017, 1ES 0229+200, 1ES 0347-121 and PKS 0548-322 do not appear in the 1FGL catalogue (Abdo et al. 2010), which includes all the sources detected above 100 MeV with a signiÞcance exceeding 4.5$\sigma $ during the first 11 months survey of LAT onboard {\it Fermi} (Atwood et al. 2009).

We used the the publicly available  data\footnote{accessible from \tt http://fermi.gsfc.nasa.gov} to search for detections or derive updated upper limits. 
We selected the photons of class 3 (DIFFUSE) with energy in the range 0.1--100 GeV collected from 2008 August 4 (MJD 54682) to 2010 July 15  (MJD 55392), for a total of about 24 months of elapsed time.  These data were processed by using Science Tools 9.15.2, which includes the Galactic diffuse and isotropic background and the Instrument Response Function IRF P6 V3 DIFFUSE. 

We selected photons in the good--time intervals and within a region of interest (ROI) with radius of 10$^{\circ}$ from the source radio position applying a cut on the zenith angle parameter ($<105^{\circ}$) to avoid the Earth albedo. The following steps are to calculate the live--time, the exposure map and the diffuse response.
 
With all these information at hands, we performed an analysis by using an unbinned likelihood algorithm ({\tt gtlike}) in three separate energy bands, namely 0.1--1, 1--10 and 10-100 GeV. The model included the isotropic and Galactic diffuse backgrounds, the source of interest, all the 1FGL sources in the ROI and, possibly, additional sources not included there but identified in the map. For all the point sources we assumed a power law spectrum, with flux and photon index as a free parameter and calculated the corresponding test statistic ($TS$, see Mattox et al. 1996 for a definition; in practice one assumes $\sqrt{TS}\simeq \sigma$, the significance of the detection). For each energy bin showing a significance smaller than $TS=16$ we derive upper limits.

For 1ES 0229+200 and 1ES 0347-121 we only obtained upper limits. For PKS 0548-322 we obtained a detection in the high energy (10-100 GeV) bin, upper limits in the other two bands.

\begin{table}
\begin{center}  
\begin{tabular}{lcc} 
\hline  
  Energy band [GeV]& $N$ & $TS$ \\
\hline 
0.1--1& 177&17\\
1--10& 45&56\\
10--100& 5&43\\ 
\hline \\
\end{tabular}
\end{center}
\caption{Results of the analysis of the LAT data for RGB J0152+017. For each energy band we report $N$, the number of photons predicted by the model, and the corresponding value of the test statistics, $TS$}
\label{lat0152}
\end{table}

For RGB J0152+017 we have detections in all three energy bands (see Table \ref{lat0152}). In this case a fit on the whole 0.1--100 GeV energy band gives a spectral slope of $\Gamma=2.0\pm 0.2$ consistent with the binned analysis described above. Since the significance of the detection is relatively large, we also derive a light-curve with a bin-size of 10 days to study possible long-term variability. The source is barely detected (with $TS>10$) in about 10 bins. For the two time bins (MJD 54872--54882 and MJD 55272--55282) showing the highest significance, $TS>12$ ($\approx 3.5 \sigma$), we also derived a spectrum, which is very different in the two cases ($\Gamma=1.15\pm0.2$ and $\Gamma=2.6\pm 0.4$, respectively). 
Although the significance of these two spectra is rather low, their difference may suggest that the long-term flat spectrum is in fact the average over hard and soft spectra. 

In the analysis of the RGB J0152+017 data we discover significative ($TS=49$) $\gamma$--ray emission from a source distant $\sim 1^{\circ}$ from RGB J0152+017. The position, R.A. $=27.15^{\circ}$, Dec.$= + 1.45^{\circ}$ (with a positional uncertainty, measured by the 95\% containment radius without systematics, of 0.06$^{\circ}$)  
corresponds to that of the flat spectrum CRATES radio source J0148+0129, associated to the source PMN J0148+0129 (of unknown redshift), not included in the 1FGL catalogue.

Note that the reprocessed emission component is in principle expected to be extended (pair ``echo"), while the procedure discussed above to derive upper limits and the fluxes assumes a point-like source or a source with an extension contained within the LAT PSF. As we verified a {\it posteriori} (see \S \ref{extension}) this is a reasonable assumption for the cases discussed in the present work.

\subsection{Multifrequency data}

SEDs are shown in Fig. \ref{sed}. We use the multifrequency data already considered in T10a and the H.E.S.S. spectrum of 1ES 0548--322 recently published by Aharonian et al. (2010). 

All the TeV spectra have been obtained by H.E.S.S. In three cases the sources have shown very small spectral changes during observations extended over several months (1ES 0347-121) or years (1ES 0229+200, PKS 0548-322 ). Based on this, as in T10a and T10b, we assume that the averaged H.E.S.S. spectra are representative of the TeV component also during the period of the LAT observations. 

For RGB J0152+017, instead, the data have been obtained in a relatively short exposure ($\sim 2$ weeks) during a ToO observation in 2007 (Aharonian et al. 2008),  and therefore the spectrum cannot be considered as representative of an average state.
This fact, together with the pronounced GeV variability suggested by LAT data (we report in Fig. \ref{sed} the two spectra discussed in \S 2.1), is probably the reason for the  apparent disagreement between the LAT and H.E.S.S. spectra visible in Fig. \ref{sed}. Considering this {\it caveat},  for completeness we still use the average LAT spectrum and the single epoch H.E.S.S. spectrum in the following analysis although the inferences that we can draw about the blazar emission model and IGMF constraints from this source will reflect the corresponding uncertainties affecting the data.

We show the most recent optical-UV and X-ray data from {\it Swift}, taken during or in vicinity of the first three months (August-October 2008) of LAT pointings. We refer the reader to T10a for the details of the analysis of these data and a general discussion of the SEDs. For RGB J0152+017 there are no {\it Swift} observations close in time to the LAT ones. As in T10a we use in the modeling the X-ray spectrum close in time to the TeV observations taken with {\it XMM-Newton} and {\it RXTE} (Aharonian et al. 2008).

In three cases (RGB J0152+017, 1ES 0229+200 and 1ES 0548--322) the optical-UV data describe a rather steep continuum that, as already discussed in Tavecchio et al. (2009) for 1ES 0229+200, traces the emission of the host galaxy. In these cases, in order not to overestimate the UV flux, it is necessary to assume a strong roll-off of the non-thermal continuum of the jet below the soft X-ray band. The optical to X-ray flux is very hard and in the most extreme case of 1ES 0229+200 requires a spectrum $F(\nu)\propto \nu^{1/3}$, as expected from the tail of the synchrotron emission of high-energy electrons.  

\section{Modeling the SED}

\subsection{Primary blazar emission model}

SEDs of TeV BL Lacs are generally adequately reproduced by one-zone leptonic models (e.g. Ghisellini et al. 1998, Tavecchio et al. 1998). Support to the idea that the emission mainly originates in a single region comes from the existence of a characteristic timescales, thought to be related to the size of the emitting region\footnote{but there are exceptions, e.g. Mkn 501 (Albert et al. 2007) and PKS 2155-304 (Aharonian et al. 2007) showing event with very short timescales.}, and from the tight correlation between variations at very different frequencies (e.g. Fossati et al. 2008)\footnote{exceptions are the so-called orphan flares (e.g. Krawczynski et al. 2004).}. 

We use the one-zone leptonic emission model fully described in Maraschi \& Tavecchio (2003). Briefly, the emitting region is a sphere with radius $R$  with a tangled and uniform magnetic field $B$. Relativistic electrons are assumed to have an isotropic distribution and to follow a smooth broken power law energy spectrum with normalization $K$ and indices $n_1$ from $\gamma_{\rm min}$ to $\gamma_{\rm b}$ and $n_2$ above the break up to $\gamma_{\rm max}$. These electrons emit through synchrotron and synchrotron-self Compton (SSC) mechanisms. SSC emission is calculated assuming the full Klein-Nishina cross section. This is particularly important for HBLs as those considered here, in which the effects of the KN cross section essentially shape the SSC component above few tens of GeV (e.g. Tavecchio et al. 1998). Relativistic amplification of the emission is described by the Doppler factor, $\delta$. These nine parameters ($R$, $B$, $K$, $n_1$, $n_2$, $\gamma _{\rm min}$, $\gamma _{\rm b}$, $\gamma _{\rm max}$, $\delta$) fully specify the model.

Once the intrinsic spectrum $F_{\rm int}(E)$ is derived, the corresponding observed spectrum is calculated as $F_{\rm int, obs}(E)=F_{\rm int}(E) e^{-\tau(E)}$, where $\tau(E)$ is the energy-dependent optical depth (specified below) for the $\gamma \gamma \rightarrow e^{\pm}$ interaction with the optical-IR photons of the extragalactic background light (EBL). 

\begin{figure*}
\psfig{file=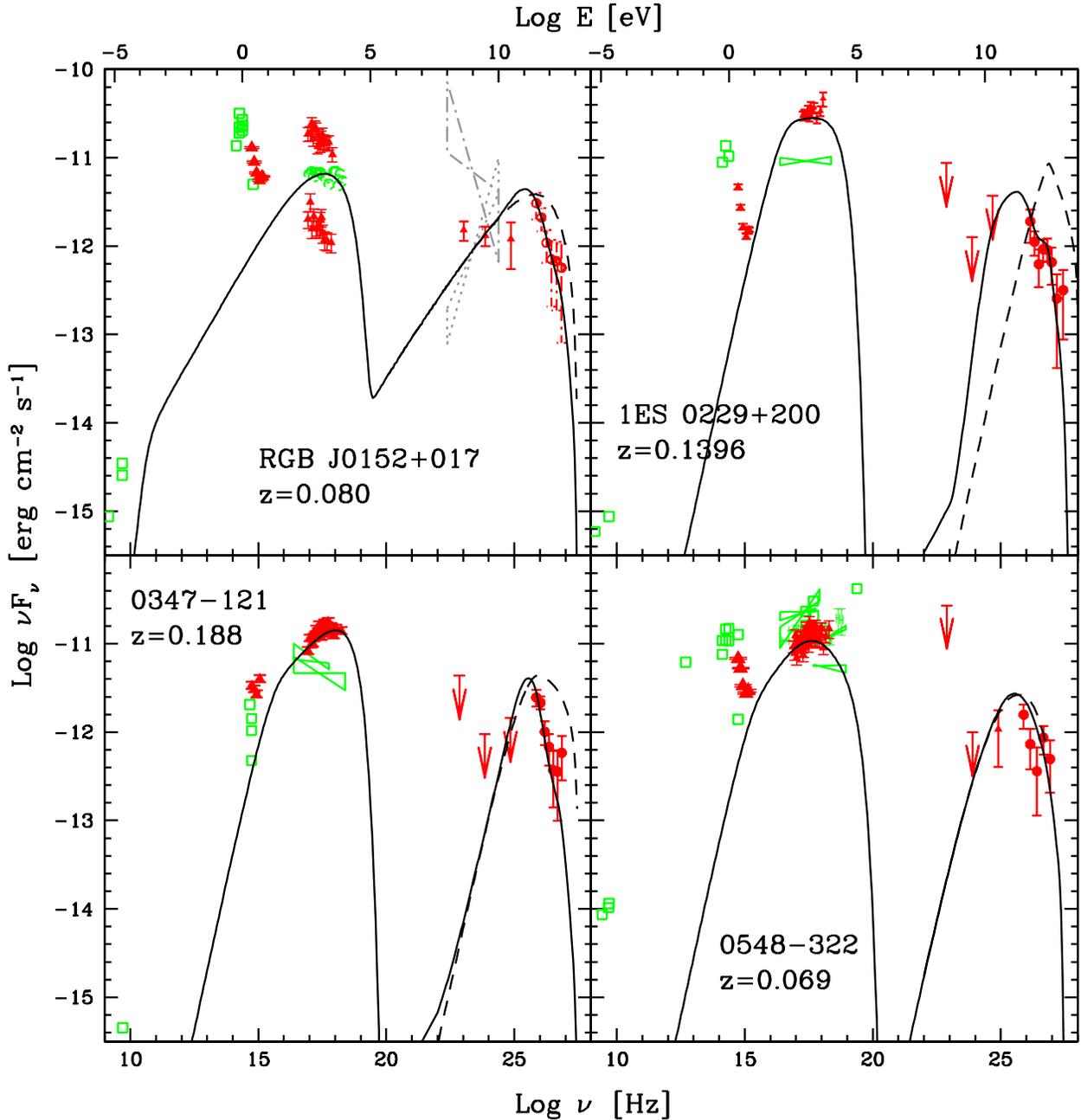,height=17.5cm,width=17.5cm}  
\caption{Spectral Energy Distributions of the sources discussed in this work. Open squares are historical data (green in the electronic version), filled triangles reports {\it Swift}/UVOT and XRT data and filled circles (red) shows the TeV spectra. The open (green) circles in the SED of RGB J0152+017 report the X-ray spectrum taken in the vicinity of the H.E.S.S. detection. {\it Fermi}/LAT data are those derived by our analysis. For RGB J0152+017 we also show (gray bow-ties) two spectra accumulated over ten days (54872-54882, dotted; 55272-55282, dashed-dotted) showing huge variations both in flux and slope (see text). See Tavecchio et al. (2010a) for references. The TeV spectrum of 1ES 0548--322 is from Aharonian et al. (2010). The solid line shows the emission model reproducing the observed SED, including synchrotron, SSC and reprocessed emission. The dashed line reports the intrinsic SSC component. See text for details.}
\label{sed}
\end{figure*}

\subsection{The reprocessed spectrum}

\subsubsection{Assumptions}

During their travel to the Earth, a fraction of the blazar primary high-energy $\gamma$--rays interacts with the low energy (optical-IR) extragalactic photons producing electron-positron pairs.
The pairs, in turn, inverse-Compton scatter the photons of the CMB, emitting a reprocessed high energy component. The flux and the spectral shape of this component depend on the spectrum of the primary $\gamma$--rays and on the intensity 
and geometry
of the intergalactic magnetic field which, deviating the pairs during their cooling, affects in an energy-dependent way the observed flux.  The comparison of the GeV data with the expected reprocessed spectrum provides a way to constrain the IGMF. We take into account these effects refining the simplified treatment of T10b.

The interaction of a  primary photon with energy (measured in TeV) $E_{\rm TeV}$   
with a low frequency (optical--IR) photon of the (EBL) results in an electron positron pair with Lorentz factor 
$\gamma \simeq E/2m_e c^2\simeq 10^6 E_{\rm TeV}$. These pairs, in turn, will inverse Compton scatter the photons 
of the CMB producing $\gamma$--rays of energy $\epsilon \simeq \gamma^2 h \nu_{\rm CMB} \simeq  2.8\, kT_{\rm CMB}\, \gamma^2 = 0.63\, E_{\rm TeV}^2$ GeV.

In T10b we described the reprocessed $\gamma$--ray spectrum as a power-law (with slope 0.5 in energy) with an abrupt cut-off above an energy $\epsilon _{\rm max}$. This approximation was motivated by the fact that this component derives from electrons which have a very short cooling timescale and therefore quickly form a completely cooled distribution. Although this approximation was appropriate for the purposes of T10b, here we adopt a more self-consistent approach, in order to have a better description of the reprocessed spectrum around $\epsilon _{\rm max}$.

Another important aspect to consider is the possibility to have more than one generation of pairs. In T10b we assumed that the maximum energy of the first generation reprocessed photons, $\epsilon _{\rm max}$, was below the threshold for further interaction with the EBL, which for sources at $z\sim 0.1$ corresponds to $\epsilon _{\rm max} \sim 300-400$ GeV. This translates into a maximum energy of the intrinsic blazar emission, $E_{\rm max}\simeq 20$ TeV. When realistic SSC spectra are considered this condition could be violated, since there could be an important tail of the intrinsic SSC component above this energy. Therefore it is necessary to consider that also the $\gamma$--ray photons produced through reprocessing can be further absorbed by the interaction with the EBL, producing a second generation of inverse Compton emitting pairs. Again, this second component could be absorbed and so on, possibly leading to the development an electromagnetic cascade. 

A complete, self-consistent treatment of the cascades could be done with Monte Carlo simulations (e.g. Coppi \& Aharonian 1997, Elyiv et al. 2009, Dolag et al. 2009, 2011, Neronov et al. 2010, Taylor et al. 2011). However for sources at the redshift around 0.1, as those considered here, a real electromagnetic cascade (i.e. more than two generations) would develop only if the maximum energy of the primary SSC radiation is very large, $E_{\rm max}>100$ TeV.
This can be derived from the relation between the energy of the primary, $E_{\rm TeV}$ (measured in TeV) and of the reprocessed $\epsilon $ photon 
$\epsilon \simeq 0.63 E_{\rm TeV}^2$ GeV. We make the simplified assumption that the absorption of the radiation produced by pairs becomes important only if its maximum energy is above a threshold $\epsilon _{\rm th}$ at which the optical depth is larger than one. Therefore an important second generation of pairs will develop only if the maximum energy of the photons produced by the first-generation satisfies $\epsilon _{\rm max}=0.63 E_{\rm max}^2>\epsilon _{\rm th}$, namely, if the maximum energy of the primary photons exceeds: $E_{\rm max}>(\epsilon _{\rm th}/0.63)^{1/2}$ (again, $E$ is in TeV, $\epsilon$ in GeV). 
Generalizing this expression, the condition for the effective absorption of the radiation of the $n$-th generation of pairs (and therefore for the production of an important population of pairs of the generation $n+1$) is:
\begin{equation}
E_{\rm max}>1600 \left( \frac{\epsilon_{\rm th}}{1.6\times10^6} \right) ^{\frac{1}{2n}}  \,\,\, {\rm TeV}
\label{ngen}
\end{equation}
\noindent
Assuming $\epsilon _{\rm th}=500-1000$ GeV, a value suitable to describe the case of the sources analized here (see below), to have more than two generations, $n=2$ in Eq. \ref{ngen}, $E_{\rm max}$ must be larger than $\approx 250$ TeV (due to the small exponent this value is not strongly dependent on $\epsilon _{\rm th}$, the assumed threshold for important absorption). In the context of the standard SSC model used below such large energies are very difficult to achieve: therefore we can limit the calculation of the reprocessed spectrum to the second generation.

A further assumption concerns the variability of the primary TeV emission. As noted by Dermer et al. (2010), the reprocessed radiation in the GeV range reaches the observer with delays with respect to the direct unabsorbed radiation. With magnetic fields of the order of $10^{-15}$ G such as those derived below, the delays are of the order of some million years. Therefore, to compare the flux of the reprocessed component with the observer primary flux we have to assume that the primary $\gamma$--ray flux is on average stable over times of the order of $10^7$ years. 
Smaller times would result in correspondingly lower values for the IGMF. For the case of 1ES 0229+200, Dermer et al. (2010), considering a lower limit of $t>1$ year for the activity timescale of the source find $B>3\times 10^{-19}$ G.

In the calculation the use of a specific model for the EBL is unavoidable. In the range of wavelength interesting here, $0.1-15$ $\mu$m, several of recent EBL estimates converge: the recent models of Franceschini et al. (2008), Gilmore et al. (2009), Finke et al. (2010), Dominguez et al. (2010), agree well (but see Orr et al. 2011) and are consistent with the low level of the EBL suggested by the recent observations of Cherenkov telescopes (e.g. Aharonian et al. 2006, Mazin \& Raue 2007; see also Kneiske \& Dole 2010). We adopt the {\it LowSFR} model of Kneiske et al. (2004) which provides an optical depth similar to that of all the other updated models up to energies of 4--5 TeV. For larger energies the predicted optical depth is {\it lower} than that of the other models. In this respect in our calculation the importance of the reprocessed component is minimized and thus our results can be considered as conservative.

Finally, we note that in our derivation we are implicitly assuming that the magnetic field is oriented perpendicularly to the direction of the relativistic pairs 
and maintains its coherence within the region in which pairs emit and cool. A non perpendicular magnetic field is less effective in deviating the emitting pairs, therefore determining a higher final observed flux.
On the other hand, if the coherence length $\lambda _B$ of the field (namely the size of the regions in which the field maintain its coherence) is smaller than the typical cooling length of the pairs $c t_{\rm cool}\simeq 0.7 \gamma _6^{-1}$ Mpc, the resulting magnetic field will be correspondingly larger, being in this case $B\propto \lambda_B ^{-1/2}$ (e.g. Neronov \& Semikoz 2009). In any case, both effects would result in a larger value for the IGMF intensity making our lower limits {\it conservative} estimates.

\subsubsection{First generation spectrum}
\label{firstgen}

We calculate the energy distribution of the emitting pairs assuming the equilibrium (ensured by the very small cooling time of the pairs) between injection of new energetic pairs in the volume between the source and the Earth (produced through the conversion of absorbed $\gamma$--rays) and their cooling through IC emission. From the general kinetic equation for the electron distribution $N(\gamma)$ we get:
\begin{equation}
N(\gamma)=\frac{\int _{\gamma} ^{\infty} Q(\gamma)}{\dot{\gamma}}
\label{ngamma}
\end{equation}
\noindent
where the injection term $Q(\gamma)$ is calculated considering the absorption process:
\begin{equation}
Q(\gamma) = k \, \frac{F_{\rm int}(E)}{E}\left[1-e^{-\tau(E)} \right]     \;\;\;\; {\rm with} \;\;\;\ E=2m_ec^2 \gamma
\label{qgamma}
\end{equation} 
\noindent
where $F_{\rm int}(E)$ is the spectrum of the intrinsic (primary) blazar emission and $\tau(E)$ is the optical depth for absorption of $\gamma$--rays.
The cooling term $\dot{\gamma}$ in Eq. \ref{ngamma} is specified assuming that the cooling is dominated by the IC scattering on the CMB photons in the Thomson limit:
\begin{equation}
\dot{\gamma}\simeq \frac{4}{3} \frac{\sigma _T}{m_{\rm e}c} U_{\rm CMB} \gamma ^2.
\end{equation} 
where $U_{\rm CMB}$ is the CMB energy density at the redshift of the reprocessing region.

\begin{figure}
\psfig{file=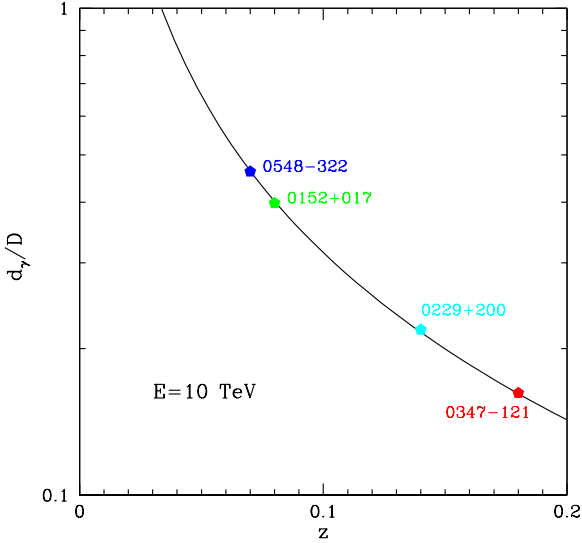,height=8.5cm,width=8.5cm}  
\caption{Ratio of the $\gamma$--ray mean free path for photons of $E=10$ TeV and source distance versus redshift 
 using the {\it LowSFR} model of Kneiske et al. (2004). Stars indicate the redshift of the sources considered in our work.}
\label{lambda}
\end{figure}

\begin{table*}
\begin{center}  
\begin{tabular}{lcccccccccc} 
\hline  
     &$\gamma _{\rm min}$ &$ \gamma_{\rm b}$ & $\gamma _{\rm max}$& $n_1$&  $n_2$ & $B$ [G]& $K$ [cm$^{-3}$]& $R$ [cm]& $\delta $ & $B_{\rm IGMF}$ [G]\\
\hline 
RGB 0152+017  & 10 & $3\times 10^5$& $10^6$& 2& 3.5 & $3\times 10^{-2}$& $10^4$& $4\times 10^{15}$&25 &$5\times 10^{-15}$ \\
1ES 0229+200 (A) & $2.5\times 10^5$& -- & $3.5\times 10^6$& 2.85& -- & $3\times 10^{-3}$& $10^8$& $5.4\times 10^{16}$&30  &$2\times 10^{-15}$\\
1ES 0229+200 (B)  & $7.5\times 10^5$& -- & $10^7$& 2.85& -- & $7.2\times 10^{-4}$& $7.3\times 10^8$& $5.8\times 10^{16}$&34  &$7\times 10^{-14}$\\
1ES 0347--121   & $5\times 10^4$ & $1.5\times 10^6$ &$2\times 10^6$& 2.2 & 3.1 & $2\times 10^{-2}$& $1.8\times 10^4$& $3.35\times 10^{16}$&15 &$ 10^{-14}$ \\
PKS 0548--322    & $1.5\times 10^4$ &$2.8\times 10^5$  &$1.5\times 10^6$& 2.0 &4.1 & 0.1& $10^4$& $3.6\times 10^{15}$&16 &$5\times 10^{-15}$ \\
\hline \\
\end{tabular}
\end{center}
\caption{Input parameters for the emission models shown in Fig.\ref{sed}. See text for definitions. For 1ES 0229+200 we assume a simple power law electron energy distribution (see Tavecchio et al. 2009) and we also report (B) the parameters used for the model reproduced in Fig. \ref{0229ver2} and discussed in the text.}
\label{model}
\end{table*}

Once the pair energy distribution is specified through Eq. \ref{ngamma} we calculate the corresponding inverse Compton spectrum $F_1(\epsilon)$. The normalization of the spectrum is derived considering that, as said above, the system is in the so-called {\it fast cooling} regime, i.e. the total (i.e. energy integrated) emitted reprocessed power is equal to the total power injected into pairs (neglecting the rest mass of pairs), in turn equal to the energy integrated power of the absorbed radiation (e.g. T10b).

Note that since the primary blazar spectrum $ F_{\rm int}(E)$ is usually hard in the range of energies where absorption is important and $\tau(E)$ increases rapidly with energy, the spectrum of injected electrons will also be hard and therefore, consistently with the assumption of T10b, the electron energy distribution resulting from Eq. (\ref{ngamma}) will be a simple power law $N(\gamma)\propto \gamma ^{-2}$ (since the integral is almost independent on the lower limit $\gamma$), except for the highest energies, where the distribution will have a rapidly decreasing tail. 
Importantly, this also implies that the number of pairs from the conversion of the primary photons at the highest energy, $E_{\rm max}$, exceeds the number of pairs deriving from photons af lower energies. Therefore, the reprocessed spectrum is dominated by the emission from the pairs originated by the absorption of the primary photons at $E_{\rm max}$, allowing to identify a characteristic distance from the source ($d_{\gamma}$, the mean free path for photons of energy $E_{\rm max}$), at which the entire first generation reprocessed spectrum is produced.

The effect of the IGMF is to curve the trajectory of the emitting pairs, resulting in a dilution of the reprocessed spectrum within a solid angle which is a function of the pair energy and IGMF intensity, $\Omega_{\gamma}=2\pi (1-\cos\theta _{\gamma})$.  The angle $\theta _{\rm \gamma}$ can be estimated assuming it is the angle by which the electron velocity vector changes in the cooling length $ct_{\rm cool}$:
%
$\theta _{\gamma} \sim ct_{\rm cool}/r_{\rm L} = 1.17\, B_{-15} (1+z_{\rm r})^{-4} \gamma_6^{-2}$
%
\noindent
where $r_{\rm L}=\gamma m_e c^2/(eB)\simeq 2\times 10^{24} \gamma _6 B _{-15}^{-1}$ 
cm is the Larmor radius of the electron.

In T10b we assume that for values of the energy and IGMF for which the angle of curvature due to the IGMF is within the beaming solid angle of the intrinsic blazar emission, $\Omega _{\rm c}\approx \pi \theta _{\rm c}^2$, the observed spectrum is not affected by IGMF. Conversely, if the pairs are completely isotropized during their lifetime the observed flux will be suppressed by a factor $\Omega _{\rm c}/4\pi$. A convenient approximation for the
{\it observed} first generation reprocessed spectrum, including {\it both} the absorption by the EBL and the effect of the IGMF on the emitting pairs, is calculated from $F_1(\epsilon)$ as:
\begin{equation}
F_{1, \rm obs}(\epsilon)= F_1(\epsilon) \frac{e^{-\tau_1(\epsilon)}}{\Omega _{\rm c}+\Omega_{\gamma}(\epsilon)} 
\label{finale}
\end{equation}
where, $F_1(\epsilon)$ is the IC spectrum calculated above and, again, $\Omega _{\rm c}$ is the solid angle of the primary blazar emission and $\Omega_{\gamma}(\epsilon)$ is the solid angle (dependent on the photon energy and IGMF intensity) into which the radiated energy of a pair is spread (see T10b for details). For the models reported here we assume $\theta _{\rm c}=0.1$. Smaller values of the beaming angle would result in lower fluxes for the {\it fully isotropised} reprocessed component. 

\begin{figure*}
\psfig{file=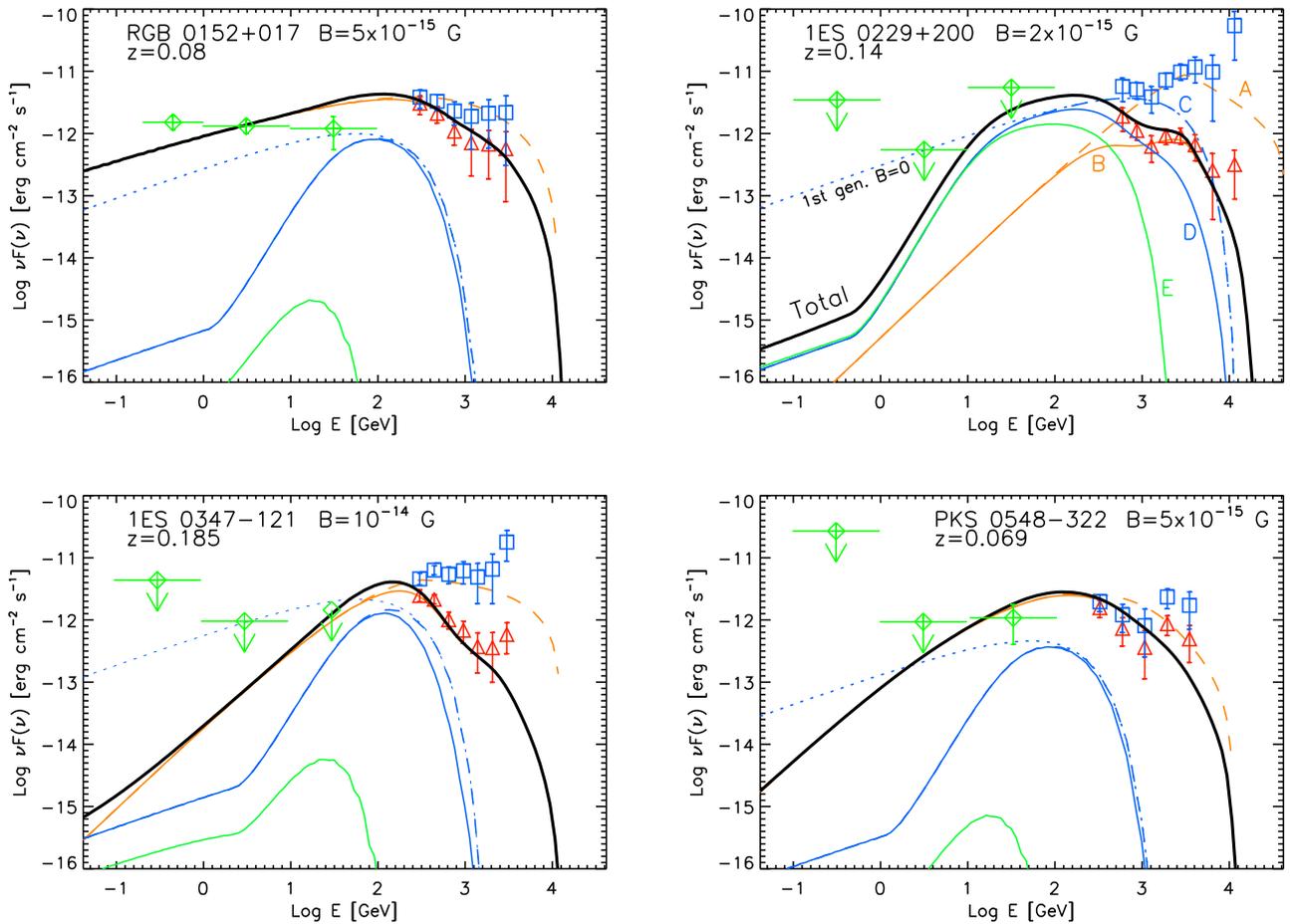,height=13.cm}
\caption{High energy SED of the sources discussed in this work. Open triangles are the observed TeV data, open squares are the data-points corrected for absorption by the EBL assuming a distance equal to that of the source. Diamonds are the LAT data-points derived by our analysis: the horizontal lines indicate the band considered to derive the flux. The solid thick line (black in the electronic version, labelled ``Total") shows the emission model reproducing the observed SED, including SSC and reprocessed emission. Thin orange lines shows the intrinsic (dashed, A) and observed (solid, B) SSC spectrum, the thin blue lines are the intrinsic (dashed, C) and observed (solid, D) first generation reprocessed spectrum. The dotted line is the first generation reprocessed spectrum assuming no IGMF. The green thin line (E) is the second generation reprocessed spectrum, providing negligible contribution in all the cases. See text for details.}
\label{sedhigh}
\end{figure*}

As long as the pairs of the first generation are produced and cool at distances from the blazar small compared to the distance of the blazar to Earth ($d_{\gamma}/D \simeq 1/\tau \ll 1$, where $d_{\gamma}$ is the mean free path of photons as measured from the blazar and $D$ the blazar distance) we can make the approximation $\tau_1\approx \tau$.
Of course the validity of this approximation depends on the considered energy and it will be violated at low energy, for which the optical depth for the primary photons is $\tau(E)<1$. However, as discussed above, we can assume that the reprocessed emission mainly derives from pairs produced by the conversion of primary $\gamma$--rays at $E_{\rm max}$, which is typically around 10 TeV. For illustration, in Fig. \ref{lambda} we report the ratio $d_{\gamma}/D$ for photons of energy $E=10$ TeV for the adopted EBL model for the range of redshift interesting for the sources considered here. For 1ES 0229+200 and 1ES 0347-121, ($z=0.14$ and 0.18, respectively) $d_{\gamma}/D \approx 0.1-0.2$, ensuring that the approximation $\tau_1\approx \tau$ is good. For the other two sources ($z=0.07$ and 0.08), instead, the approximation is not applicable. However,  for the same reason -- the optical depth is low -- 
only a small fraction of the primary continuum is reprocessed and therefore the contribution of the reprocessed emission to the total observed GeV-TeV spectrum is not important (see Fig. \ref{sedhigh}: the maximum total contribution of the reprocessed component is less than 30\%).

We finally note that the approximation considering the blazar emission pattern as a {\it uniform cone} is rather useful in simplifying the calculation and provides correct results when we are considering integrated quantities. However, the right pattern should be  considered when one is interested in the detailed spatial distribution of the reprocessed component (see \S \ref{extension}).

\subsubsection{Second generation spectrum}

The observed spectrum emitted by the second generation of pairs, $F_{2,\rm obs}({\epsilon}$), is calculated iterating the procedure described above for the first generation. 

A further assumption that we use is that the emitting pairs of the second generation are initially (i.e. when created) still collimated within the beam of the primary blazar emission. This requirement is generally satisfied since the absorbed first-generation $\gamma$--rays converting into the second generation of emitting pairs are highly energetic, $\epsilon>100-200$ GeV, and they are produced by energetic ($\gamma>10^7$), fastly-cooling pairs of the first generation. These pairs, during their short lifetime are only slightly deviated from the original direction which was that of the primary blazar emission, 
the deviation angle being $\theta < 0.6 B_{-15} \gamma ^{-2}_7$ deg. Only for values of the IGMF exceeding those considered in the following, the trajectories of the first generation pairs at the highest energy will be substantially curved during their lifetime and this approximation would break down. 

\subsubsection{Angular extension}
\label{extension}

The spreading due to IGMF implies that the reprocessed emission is not point-like but it is characterized by a finite angular size (e.g. Elyiv et al. 2009) that, especially at the higher energies (above 10 GeV), where the angular resolution of LAT is better (with containment radius of few tenths of degree), can exceed the PSF width. Under these conditions the reprocessed component should be treated as an {\it extended} source and the standard LAT analysis used above to derive upper limits or fluxes cannot be applied.

\begin{figure*}
\vskip -1.8 true cm
\hskip 1 true cm
\psfig{file=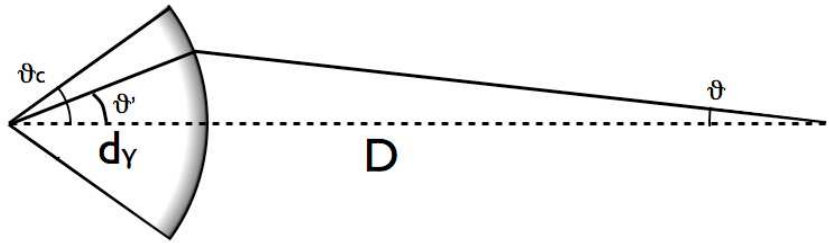,height=11.5cm}
\vspace{-2.8 truecm}
\caption{Sketch of the geometry assumed for the calculation of the spatial extension of the reprocessed emission. See text for definitions.}
\label{geometry}
\end{figure*}

It is therefore important to quantify the expected angular broadening as a function of photon energy and IGMF intensity and compare it with the LAT PSF\footnote{\tt http://www-glast.slac.stanford.edu/software/IS\\/glast$\_$lat$\_$performance.htm}.
Under the simplifying assumption that the intensity of the primary emission is uniform within the cone with semi-aperture $\theta _{\rm c}$ (see Fig. \ref{geometry}) a simple relation for the typical angular size of the reprocessed component, $\theta _{\rm v}$ can be derived (see also Neronov \& Semikoz 2009):
\begin{equation}
\theta _{\rm v}(\epsilon)\approx \min\left[ \frac{\theta _{\rm c}}{\tau -1}, \frac{\theta_{\rm \gamma}}{\tau}\right]
\label{extapprox}
\end{equation}
\noindent
where we expressed the ratio between the source distance
and the typical interaction length of photons with energy $E_{\rm max}$ (see \S \ref{firstgen}) as $\tau \approx D/d_{\gamma}>1$ (see Fig.  \ref{geometry}). Eq. \ref{extapprox} simply states that for energies and IGMF intensity for which the pair deviation is larger than the cone semi-aperture we can see the entire conversion surface, that is the spherical shell described by the angle $\theta _{\rm c}$ projected into the sky. On the contrary, when the deflection is less than $\theta _{\rm c}$, we only see the (energy dependent) portion of the surface depending by the angle $\theta _{\rm \gamma}$.

This simple estimate is derived under the assumption that the blazar uniformly illuminates the associated beaming cone. However this is a crude approximation:  indeed the blazar relativistically boosted intrinsic emission is strongly anisotropic, being concentrated along the jet axis.
To properly estimate the expected brightness profile of the reprocessed emission one should then consider in detail the emission pattern, 
that for the integrated flux is described by $\delta^4$, where $\delta$ is the beaming Doppler factor.
Due to the strong angular dependence of $\delta$ the emission pattern of the intrinsic emission is rather ``narrow" and
the observed extension size will be in general less than what derived through Eq.\ref{extapprox}.
In the following we will derive the expected profile under the simplifying assumption that the observer is exactly on-axis. More general cases will be considered elsewhere.

The flux of the reprocessed component as measured by the observer from an annulus between $\theta$ and $\theta + d\theta$ can be written as:
\begin{equation}
f(\epsilon, \theta) d\theta = 2\pi \sin \theta I(\epsilon, \theta) d\theta 
\end{equation}
\noindent
for angles $\theta<\theta _{\rm \gamma}(\epsilon)/\tau$ (for larger angles the observer does not receive flux since the pairs are not deflected enough to point toward the observer). The problem reduces then to calculate $I(\epsilon, \theta)$, the spectrum of the reprocessed component at a given observation angle. Analogously to the procedure discussed above (for an observed on axis), $I(\epsilon, \theta)$, is derived from the corresponding primary intensity at the angle $\theta ^{\prime}$, $I(E,\theta ^{\prime})$.  The relation between the two angles, $\theta ^{\prime}$ (as measured from the source) and $\theta$ (as measured form the observer) can be derived by geometrical arguments, $\theta ^{\prime}\approx (\tau-1) \theta$ assuming, as discussed above, that the reprocessed emission originates at a typical distance $d_{\gamma}$. In turn, $I(E,\theta ^{\prime})$ is calculated from that measured on-axis, $I(E,0)$ using the standard beaming relations.

Having derived $f(\epsilon, \theta )$, to evaluate the spatial size of the extended halo we calculated the (energy dependent) angle within which one collects half of the entire flux, $\theta_{0.5}(\epsilon)$ according to the following condition:

\begin{equation}
\int_{0}^{\theta _{0.5}(\epsilon)} f(\epsilon, \theta ) d\theta = 0.5 \int_{0}^{\infty} f(\epsilon, \theta ) d\theta
\end{equation}

\begin{figure*}
\vskip -0.4 true cm
\psfig{file=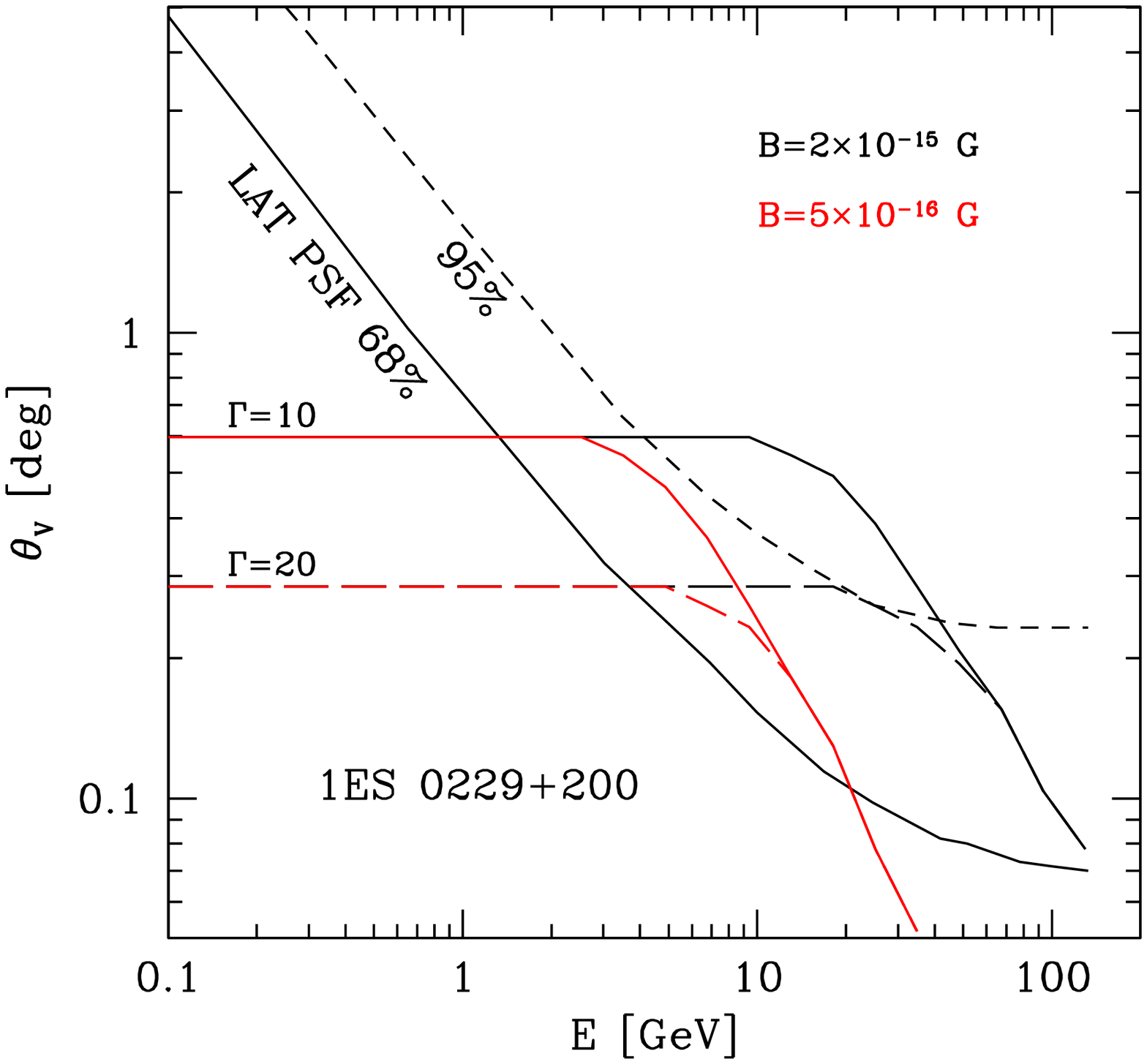,height=8.5cm}
\vskip -8.55 true cm
\hskip 7.5 true cm
\psfig{file=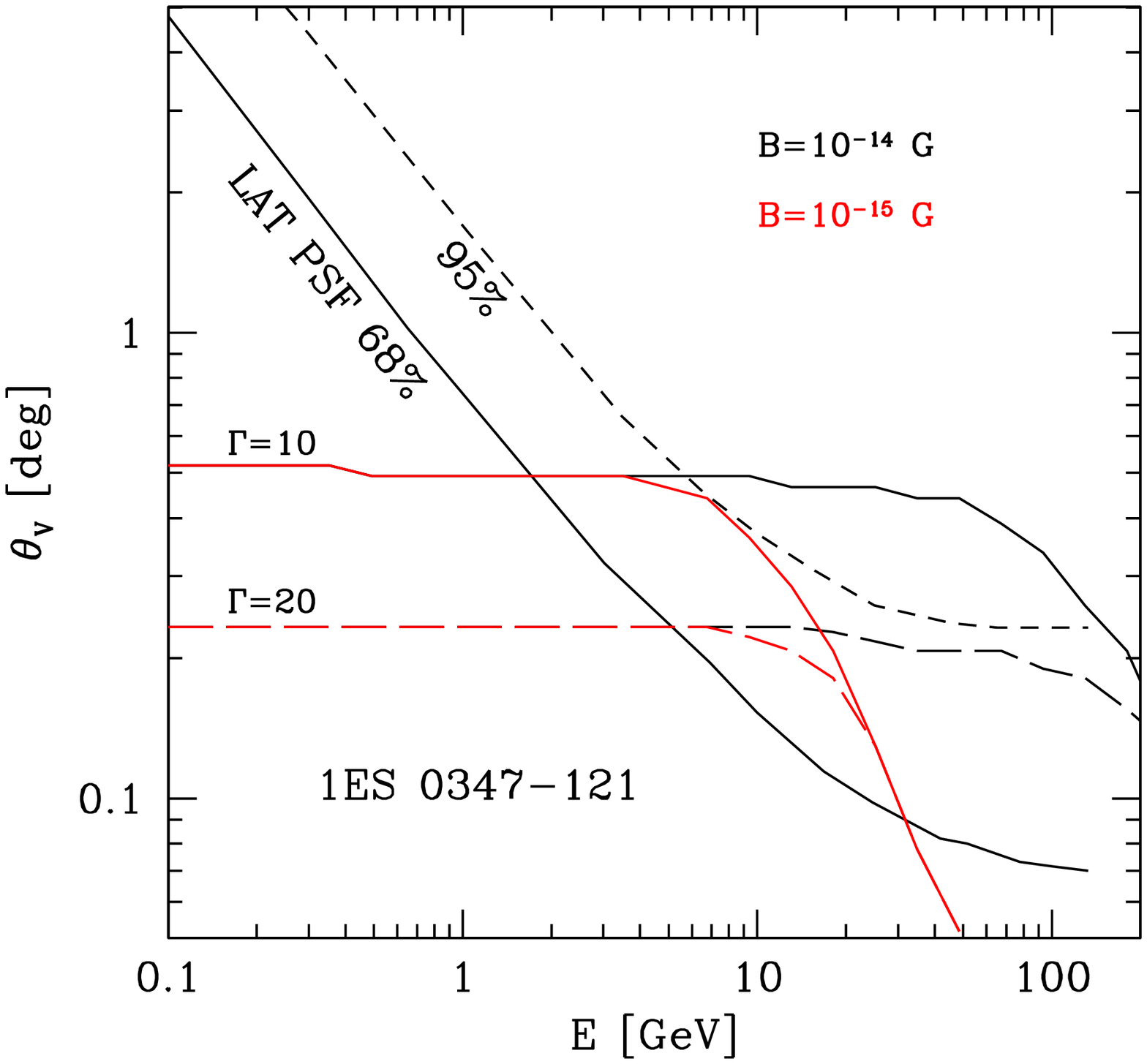,height=8.5cm}
\caption{Expected size of the reprocessed emission (expressed as the angle containing half of the total flux, $\theta _{\,\, 0.5}$) as a function of photon energy for 1ES 0229+200 and 1ES0347-121 for different values of the blazar bulk Lorentz factor $\Gamma$ and IGMF intensity. For comparison also the 68\% and 95\& LAT containment radius are reported.}
\label{ext}
\end{figure*}

Curves reporting $\theta _{0.5}(\epsilon)$ for different values of $\Gamma$ and $B_{\rm IGMF}$ for 1ES 0229+200 and 1ES 0347-121 are compared with  the LAT containment radius at 68\% and 95\% in Fig. \ref{ext} (we do not consider the other two sources since, in our models, the GeV flux is dominated by the point-like blazar emission). 
Not surprisingly, the curves display the same behavior of the simple estimate given by Eq. \ref{extapprox}. At low energy $\theta _{0.5}$ does not depend on the energy because the emission of the corresponding electrons is isotropic and the extension of the reprocessed emission is dictated by the typical beaming angle of the primary continuum, $\approx 1/\Gamma$. At sufficiently large energy, instead,  the electrons start to emit within small angles and thus the extension is fixed by the corresponding deviation angle of the electrons, $\theta _{\gamma}\propto \epsilon ^{-1}$.
The values of $\theta _{0.5}$ are always within the containment radius of LAT and therefore we can safely use the standard analysis to derive the upper limits.


\section{Results}

The models used to reproduce the observed SEDs are reported in Fig. \ref{sed} and the input parameters are listed in Table 1. As detailed in \S 3.2, in reproducing the high-energy part of the SEDs we consider the contribution of the absorbed intrinsic SSC component, that of the absorbed first generation reprocessed component and we also consider the second-generation emission (although providing a negligible contribution in most cases). The latter two contributions depend on the assumed SSC spectrum and on the value of the IGMF intensity, $B_{\rm IGMF}$ (also reported in Tab. 1). For simplicity, in the SEDs we only report the intrinsic SSC component (dashed line) and the total (intrinsic plus reprocessed) observed spectrum (solid line). A zoom on the high-energy band, reporting  all the different components, is reported in Fig. \ref{sedhigh}, showing the intrinsic SSC spectrum (orange), the first generation reprocessed spectrum (blue) and the second generation spectrum (red). In all the cases the dashed lines report the emitted spectrum, the solid line the spectrum after absorption by interaction with the EBL. The dotted line correspond to the first-generation reprocessed spectrum in absence of any IGMF. The black thick solid line is the total observed spectrum. For comparison in Fig. \ref{sedhigh} we report both the observed TeV data and the data deabsorbed assuming the Kneiske et al. (2004) {\it LowSFR} model.

Even if possibly overestimated with our assumptions, the contribution of the second-generation component is almost negligible in all the cases (but for 1ES 0229+200). Conversely, the first generation component is generally important, especially at energies around few hundreds GeV. The importance of the contribution of this component mainly depends on the shape and maximum energy of the intrinsic SSC emission: a hard spectrum peaking at large energies (10 TeV) will result in a large absorbed power and, consequently, a large power in the reprocessed component which extends in the TeV band. On the other hand, if the primary SSC spectrum peaks below 10 TeV the reprocessed component will contribute mainly at sub TeV energies. 
Generally for values of the IGMF above $B_{\rm IGMF}=10^{-15}$ G the contribution of the reprocessed emission in the 0.1-100 GeV band goes below the present sensitivity of LAT.

The most stringent lower limit to the IGMF intensity is provided by 1ES 0347--121 (that located at the largest distance): in order not to exceed the upper limit in the 10-100 GeV band one has to use an IGMF intensity of $B_{\rm IGMF}=10^{-14}$ G. For 1ES 0229+200 the same energy bin constrains $B_{\rm IGMF}=2\times 10^{-15}$ G (comparable to that derived in T10b). For the other two sources, due to the smaller distance the contribution of the reprocessed component is less important and basically the value of $B$ is unconstrained.

\begin{figure}
\psfig{file=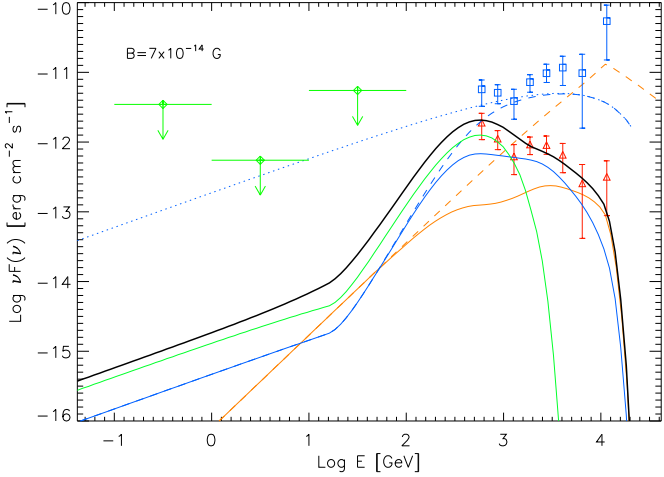,height=6.3cm}
\caption{High-energy SEDs of 1ES 0229+200 reproduced with a model assuming a large peak energy (10 TeV) for the primary SSC component. The consequent high level of the reprocessed emission (dominating the overall emission at TeV energies) requires a large value of the IGMF to not exceed the observational points below 5 TeV.}
\label{0229ver2}
\end{figure}

Since the level of the reprocessed component depends on the primary SSC spectrum, also the limits on the IGMF based on the contribution of this component to the overall spectrum will depend on it.
As an example of the importance of the intrinsic SSC spectrum in shaping the reprocessed component and of the uncertainties in the derived parameters we report in Fig. \ref{0229ver2} an alternative fit for 1ES 0229+200 assuming a SSC component peaking at 10 TeV. The first generation reprocessed component extends well in the TeV band, dominating the spectrum below 5 TeV. Also the second generation component is now very powerful
with a flux of the same order of that of the first generation component.
In order not to exceed the level of the data around 200 GeV one is forced to assume a very large IGMF, $B=7\times 10^{-13}$ G. We stress that in this case the limit to the IGMF intensity is constrained by the data-points at 100-200 GeV, not by the LAT upper limits.

This example makes clear how the derivation of the intrinsic spectrum of the source (and hence of the physical parameters of the emitting region) and the estimate of the IGMF are interlaced. In particular it makes clear that some of the models presented in T10a (in particular for the cases of 1ES 0229+200 and 1ES 0347--121) are no longer adequate to reproduce the data once the reprocessed component is taken into account, unless one assumes large values for the IGMF. As in T10b, here we prefer to adopt a conservative approach, trying to minimize the value of the required value of the IGMF. Therefore, the model reported in Figs. \ref{sed} and \ref{sedhigh} assumes a SSC peak at energies lower than assumed in T10a. This choice, in turn, implies that we derive value of the magnetic field in the emission region larger than that found in T10a. 

\begin{figure}
\vskip -1 true cm
\psfig{file=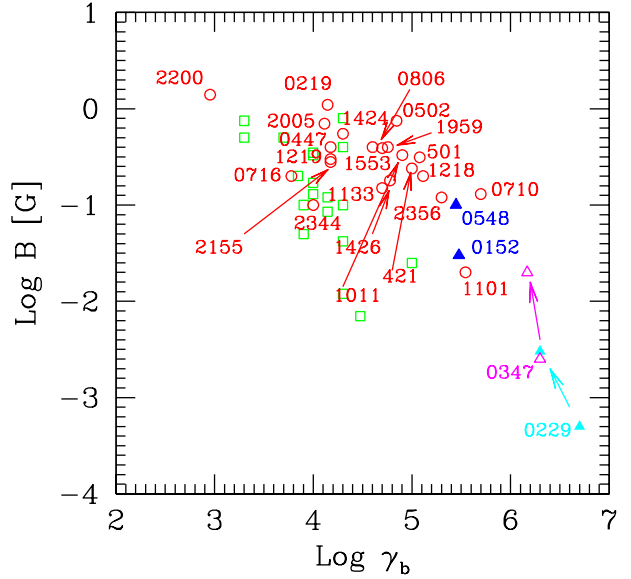,height=10cm,width=10cm}  
\vskip -0.7 true cm
\caption{Magnetic field versus the break Lorentz factor $\gamma _{\rm b}$ for the high-energy detected BL Lacs (LBAS and Cherenkov) modeled with the one-zone synchrotron-SSC model, updated from Tavecchio et al (2010b). Red open circles (triangles) show the values for the known TeV sources detected (not detected) by LAT.  Open green squares are for the LAT BL Lacs not detected in the TeV band. For 1ES 0229+200 and 1ES 0347--121 we report both the points corresponding to the parameters assumed in T10a (filled triangles) and those derived here (open triangles) considering the reprocessed component. For the other, close-by sources, the parameters adopted in T10a adequately reproduce the SED even with reprocessing.}
\label{fitparam}
\end{figure}

In Fig. \ref{fitparam} we report the updated plot from T10a showing the value of the magnetic field in the emission region and of the break Lorentz factor of the relativistic electrons (those emitting at the SED peaks) for the sample of the high-energy emitting BL Lacs considered in T10a. For the four sources considered in the present work we show both the values derived in T10a and those derived here. The different symbols indicate: sources detected both by LAT (belonging to the LBAS list Abdo et al. 2009) and Cherenkov telescopes (red open circles), by LAT only (open green squares) and by Cherenkov telescopes only (i.e. the four sources considered in this work, triangles). For the latter sources we report both the values derived in T10a (filled triangles) and those inferred here (open triangles). As discussed in T10a, the sources of the latter group populate an extreme region of the plane $B-\gamma _{\rm b}$, at very low values of the magnetic field, $B<10^{-2}$ G, and high electron energies, $\gamma_b>10^{-5}$.
As discussed above, in order to minimize the importance of reprocessed component we assume for  1ES 0229+200 and 1ES 0347--121a SSC component peaking at lower energies than those adopted in T10a. This, in turn means a larger magnetic field, as shown by the position of the new points in the plot. For RGB 0152+017 and PKS 0548-322, instead, the small redshift makes the reprocessed component less important than in the other two sources and we can model their SEDs with the same parameters of T10a.

\section{Conclusions}

The inclusion of the emission arising from the reprocessing of the absorbed primary $\gamma$ rays  can play an important role in the detailed modeling of the SEDs of blazars. In particular we have shown that for sources with hard primary VHE continuum and redshift around 0.1 the reprocessed component can even dominate below few TeV, unless the value of the IGMF intensity exceeds $10^{-13}$ G. 

The possibility that the sub-TeV radiation from blazars receives an important contribution for the reprocessing of the high-energy power absorbed through interaction of high-energy photons with the infrared-optical background has a number of important consequences. 
The first obvious fact is that the reprocessed component should be considered in the modeling. However, since the exact shape of this component is linked to the unknown value of the IGMF and the often not well determined maximum energy of the primary SSC component, this introduces several unknown parameters in the fit, breaking the simplicity of the one-zone leptonic model and preventing to obtain a unique set of physical parameters.

We would like to stress that to constrain the IGMF using the limits to the reprocessed component it is mandatory to consider realistic SSC spectra for the primary radiation. In particular, the widely considered one-zone leptonic model with typical parameters hardly predicts SSC peaks above 20-30 TeV, even in the case of the most extreme HBL objects as those considered here. This fact limits the total power that can be reprocessed into the GeV band and implies that, for the sources located at typical distance $z=0.1-0.2$, a real electromagnetic cascade does not develop and the reprocessing process ends after the second generation of pairs. 
However, blazar primary spectra could extend well above the limits derived in the SSC model discussed here if, for instance, the VHE emission is the result of hadronic processes (e.g. M{\"u}cke et al. 2003). In this case, the reprocessing could extend beyond the second generation, making the use of Monte Carlo calculations unavoidable. In this case, if the primary spectrum extends well above 10 TeV with a hard slope, the total amount of reprocessed radiation increases, implying larger values of $B_{\rm IGMF}$.

The fact that the reprocessed component can provide an important contribution even at energies in the TeV range can have a number of important consequences for the next generation of planned Cherenkov facilities (CTA\footnote{\tt http://www.cta-observatory.org/}). In particular, this fact highlights the possibility to obtain improved limits to the intensity of the IGMF by using precise measures of the overall VHE spectrum alone, especially if low thresholds can be reached.

A natural and testable consequence of the idea that the emission of blazar above 50-100 GeV is a mixture of the primary (at the highest energy) and reprocessed (at lower energies) emission is the possibility to have spectra more complex than the widely assumed simple power law,  with even multiple bumps marking the peak of the emission of different generations of pairs. Variability could also be used to disentangle different components in the $\gamma$--ray spectrum: indeed, as long as the magnetic field exceeds $B=10^{-20}$ G, the emitting pairs are effectively deviated and, consequently, the reprocessed emission is diluted over times that can reach $\approx 10^7$ years around 1-10 GeV (e.g. Neronov \& Semikoz 2009, Dermer et al. 2010). Therefore we expect that the reprocessed emission is rather stable and can be easily distinguished from the highly variable blazar primary continuum.

\section*{Acknowledgments}
We are grateful to the referee for constructive comments that helped to clarify the text.
We thank G. Ghirlanda for useful discussions.
This work was partly financially supported by a 2007 COFIN-MIUR grant.

\end{document}